\documentclass[twocolumn,aps,pra,amsmath,amssymb]{revtex4-1}
\usepackage[latin9]{inputenc}
\usepackage{amsmath}
\usepackage{amssymb}
\usepackage{graphicx}
\usepackage{esint}

\makeatletter
\usepackage{epsfig}
\usepackage{bm}
\usepackage{dcolumn}
\usepackage{color}

\makeatother

\begin{document}
\title{Fermi polarons at finite temperature: Spectral function and rf-spectroscopy}
\author{Hui Hu$^{1}$ and Xia-Ji Liu$^{1}$}
\affiliation{$^{1}$Centre for Quantum Technology Theory, Swinburne University
of Technology, Melbourne, Victoria 3122, Australia}
\date{\today}
\begin{abstract}
We present a systematic study of a mobile impurity immersed in a three-dimensional
Fermi sea of fermions at finite temperature, by using the standard
non-self-consistent many-body $T$-matrix theory that is equivalent
to a finite-temperature variational approach with the inclusion of
one-particle-hole excitation. The impurity spectral function is determined
in the real-frequency domain, avoiding any potential errors due to
the numerical analytic continuation in previous $T$-matrix calculations
and the small spectral broadening parameter used in variational calculations.
In the weak-coupling limit, we find that the quasiparticle decay rate
of both attractive and repulsive polarons does not increase significantly
with increasing temperature, and therefore Fermi polarons may remain
well-defined far above Fermi degeneracy. In contrast, near the unitary
limit with strong coupling, the decay rate of Fermi polarons rapidly
increase and the quasiparticle picture breaks down close to the Fermi
temperature. We analyze in detail the recent ejection and injection
radio-frequency (rf) spectroscopy measurements, performed at Massachusetts
Institute of Technology (MIT) and at European Laboratory for Non-Linear
Spectroscopy (LENS), respectively. We show that the momentum average
of the spectral function, which is necessary to account for the observed
rf-spectroscopy, has a sizable contribution to the width of the quasiparticle
peak in spectroscopy. As a result, the measured decay rate of Fermi
polarons could be significantly larger than the calculated quasiparticle
decay rate at zero momentum. By take this crucial contribution into
account, we find that there is a reasonable agreement between theory
and experiment for the lifetime of Fermi polarons in the strong-coupling
regime, as long as they remain well-defined. 
\end{abstract}
\maketitle

\section{Introduction}

The polaron problem that describes a single impurity interacting with
a host environment is a long-lasting research topic in modern physics
\cite{Alexandrov2010}. The initial study can be traced back to the
seminal work by Landau on the description of electron motion in crystal
lattices \cite{Landau1933}. The resulting quasiparticle picture plays
a fundamental role in understanding the complex quantum many-body
physics, occurring in solid state systems \cite{Mahan1967,Roulet1969,Nozieres1969},
helium liquids \cite{Bardeen1967} and most recently in ultracold
atomic quantum gases \cite{Massignan2014,Lan2014,Schmidt2018}. The
latest development with ultracold atoms is particularly exciting,
since highly imbalanced quantum mixtures present a clean and controllable
test-bed that is well-suited to explore the limits of Landau's quasiparticle
paradigm. In the extremely imbalanced case, the minority component
of mixtures realizes the single-impurity limit, and the interaction
between the impurity and surrounding environment (i.e., the majority
component) can be precisely tuned to be arbitrarily strong, by using
the so-called Feshbach resonance technique \cite{Bloch2008,Chin2010}.
As a result, one can systematically investigate the polaron physics
with unprecedented precision in the strong-coupling regime \cite{Massignan2014,Lan2014,Schmidt2018}.

The rapid growing interest on the ultracold atomic polaron physics
already leads to a number of breakthrough experimental discoveries
over the past fifteen years \cite{Schirotzek2009,Zhang2012,Kohstall2012,Koschorreck2012,Cetina2016,Hu2016,Jorgensen2016,Scazza2017,Zan2019,Zan2020,Ness2020},
inspired by the celebrated Chevy's variational ansatz for Fermi polarons,
which describes the dressing of the impurity motion with one particle-hole
excitation of the host Fermi sea \cite{Chevy2006}. The ground-state
attractive Fermi polarons was first realized with $^{6}$Li atoms
in 2009 at Massachusetts Institute of Technology (MIT) \cite{Schirotzek2009},
by using the ejection radio-frequency (rf) spectroscopy through the
measurement of the transferring rate of the impurity to a third, unoccupied
hyperfine state. In 2012, novel excited state of repulsive Fermi polarons
was subsequently observed \cite{Kohstall2012,Koschorreck2012}. Measurements
have also extended to Bose polarons in 2016 \cite{Hu2016,Jorgensen2016},
where the host environment is given by a weakly interacting Bose-Einstein
condensate. Those milestone experiments motivated numerous theoretical
works \cite{Combescot2007,Prokofev2008,Punk2009,Cui2010,Massignan2011,Schmidt2012,Parish2013,Vlietinck2013,Rath2013,PenaArdila2015,Levinsen2015,Goulko2016,Hu2018,Tajima2018,Tajima2019,PenaArdila2019,Mulkerin2019,Liu2019,Wang2019,Liu2020,Adlong2020,Parish2021,Pessoa2021,Hu2021}.

For Fermi polarons, most of the theoretical studies focus on the idealized
case of zero temperature. This is reasonable, since the experiments
were mostly carried out at low temperatures, where the finite-temperature
effect could be negligible. However, Fermi polarons at nonzero temperature
are also of great interest, particularly in the strong-coupling regime.
In a recent experiment at MIT \cite{Zan2019}, the temperature evolution
of the rf spectroscopy of unitary Fermi polarons with infinitely large
coupling constant right on Feshbach resonance was measured up to two
times Fermi temperature, $2T_{F}$. The breakdown of polaron quasiparticle
near Fermi degeneracy was clearly demonstrated.

On the theoretical side, finite-temperature quasiparticle properties
of Fermi polarons are less understood. The first theoretical investigation
of finite-temperature Fermi polarons by the present authors and co-workers
is restricted to the low-temperature regime (i.e., $T<0.2T_{F}$)
\cite{Hu2018}, where the quasiparticle properties such as the decay
rate are extracted from the finite-temperature Green function of the
impurity, by treating the smallest fermionic Matsubara frequency (in
absolute magnitude) $\pi k_{B}T$ as a small parameter. This restriction
can be lifted, either by directly using a retarded Green function
in the real-frequency domain \cite{Mulkerin2019}, or by applying
the analytic continuation to convert the Matsubara frequency to real
frequency \cite{Tajima2018,Tajima2019}. The latter may suffer from
some uncontrollable errors, since, strictly speaking, the numerical
analytic continuation is not a well-defined procedure \cite{Goulko2016,Haussmann2009}.
Alternatively, an interesting finite-temperature variational approach
has recently been proposed by Meera Parish and her collaborators \cite{Liu2019,Liu2020,Parish2021}.
By solving the Chevy ansatz (extended to finite temperature) at the
level of one-particle-hole excitation and keeping a sufficiently large
number of discrete eigenstates \cite{Liu2019}, both short-time dynamics
and rf-spectroscopy of Fermi polarons at finite temperature have been
investigated in detail. However, the finite-temperature quasiparticle
properties of Fermi polarons over a wide temperature regime has not
been addressed, probably due to the difficulty of reaching the continuous
limit, where an infinitely large number of truncated basis is needed.

In this work, we aim to present a systematic study of quasiparticle
properties of both attractive and repulsive Fermi polarons, by using
the conventional non-self-consistent many-body $T$-matrix theory
in the single-impurity limit \cite{Combescot2007,Massignan2011}.
The theory is fully equivalent to the finite-temperature variational
approach \cite{Liu2019,Parish2021}, but has the advantage and simplicity
of avoiding a small spectral broadening parameter, which might be
needed in the calculations of spectral function and spectroscopy.
As we work directly in the real-frequency domain, our calculations
are also free from any potential errors arising from the numerical
analytic continuation. Our main results can be briefly summarized
as follows. 

First, we present a detailed study of the impurity spectral function,
from which we extract the finite-temperature decay rate or lifetime
of Fermi polarons. We find somehow surprisingly that, with a weak
coupling strength between the impurity and the Fermi sea (i.e., $1/(k_{F}a)\leq-1$
for attractive polarons and $1/(k_{F}a)\geq1$ for repulsive polarons),
the decay rate of both attractive and repulsive Fermi polarons at
zero momentum does not increases significantly at high temperature,
indicating the existence a well-defined quasiparticle far above the
Fermi degeneracy. In sharp contrast, in the strong-coupling limit,
the lifetime of Fermi polarons rapidly increases with increasing temperature,
and in the unitary limit the quasiparticle picture of Fermi polarons
already breaks down near Fermi degeneracy. This observation agrees
well with our previous $T$-matrix results on the width of quasiparticle
peak in the ejection rf-spectroscopy \cite{Mulkerin2019}.

On the other hand, by utilizing the single-impurity limit, we can
now calculate the impurity spectral function in a more efficient way,
much faster than that in our previous work \cite{Mulkerin2019}, which
was carried out at finite impurity density and has a bottleneck in
the impurity self-energy calculation with real-frequency. The consideration
of the single-impurity limit is also more physical, as we are interested
in the single-polaron problem, and the polaron-polaron interaction
at finite density then should be irrelevant and should be avoided.
Taking the advantage of a fast calculation of the spectral function,
we are able to carefully examine the ejection and injection rf-spectroscopy.
As the rf-spectroscopies of Fermi polarons in recent experiments \cite{Scazza2017,Zan2019}
are not momentum-resolved, we pay specific attentions to the possible
effect of the momentum average on the measured width of the quasiparticle
peak in the spectra. We find that the measured width is typically
much larger than the decay rate of Fermi polarons at zero momentum,
according to our theoretical simulations.

For the finite-temperature measurement of unitary Fermi polarons at
MIT \cite{Zan2019}, we find a good agreement between theory and experiment
below a characteristic temperature $T<0.7T_{F}$, without any free
parameters. Above this temperature, the quasiparticle picture starts
to break down and our non-self-consistent $T$-matrix approach (or
finite-temperature Chevy ansatz with one-particle-hole excitation)
is not able to capture the key physics. A refined theoretical treatment
is therefore needed. For the low-temperature measurement of repulsive
Fermi polarons at European Laboratory for Non-Linear Spectroscopy
(LENS) \cite{Scazza2017}, we show that the decay rate determined
from Rabi oscillations cannot be theoretically explained solely by
considering the decay rate of zero-momentum repulsive polarons even
at nonzero temperature \cite{Adlong2020}. It can be quantitatively
understood, only when we take into account the momentum average in
the impurity spectral function.

The rest of the manuscript is organized in the following way. In Sec.
II, we briefly describe the non-self-consistent $T$-matrix approach
for a single Fermi polaron at finite temperature. In Sec. III, we
discuss in detail the impurity spectral function and the associated
quasiparticle decay rate, as functions of temperature and coupling
strength. In Sec. IV and Sec. V, we consider the ejection and injection
rf-spectroscopies, respectively. We compare our theoretical predictions
with the measurements at MIT and LENS, without adjustable parameters.
We comment briefly on how to further improve the theoretical description
of Fermi polarons, by going beyond the simple $T$-matrix approach.
In Sec. VI, we consider the cases with unequal mass between the impurity
and the host environment. In Sec. VII, we summarize the results. Finally,
in Appendix A we present some subtle details of our numerical calculations.

\section{The non-self-consistent T-matrix approach for a single impurity}

We consider an impurity of mass $m_{I}$ interacting with a homogeneous
bath of fermionic atoms of mass $m$ in three dimensions, as described
by the model Hamiltonian (the system volume is set to unity, $V=1$),

\begin{equation}
\mathcal{H}=\sum_{\mathbf{k}}\epsilon_{\mathbf{k}}c_{\mathbf{k}}^{\dagger}c_{\mathbf{k}}+\sum_{\mathbf{p}}\epsilon_{\mathbf{p}}^{(I)}d_{\mathbf{p}}^{\dagger}d_{\mathbf{p}}+g\sum_{\mathbf{kpq}}c_{\mathbf{k}}^{\dagger}d_{\mathbf{q}-\mathbf{k}}^{\dagger}d_{\mathbf{p}}c_{\mathbf{q}-\mathbf{p}},
\end{equation}
where $c_{\mathbf{k}}^{\dagger}$ and $d_{\mathbf{p}}^{\dagger}$
are the creation field operators for fermionic atoms and the impurity,
respectively. The first two terms in the Hamiltonian are the single-particle
terms with dispersion relation $\epsilon_{\mathbf{k}}=\hbar^{2}\mathbf{k}^{2}/(2m)$
and $\epsilon_{\mathbf{p}}^{(I)}=\hbar^{2}\mathbf{p}^{2}/(2m_{I})$,
while the last term describes the $s$-wave contact interaction with
a bare coupling strength $g$. It is well-known that the use of the
contact interaction potential is not physical at high energy, and
the associated ultraviolet divergence could be removed by using the
standard regularization relation, 
\begin{equation}
\frac{1}{g}=\frac{m_{r}}{2\pi\hbar^{2}a}-\sum_{\mathbf{p}}\frac{2m_{r}}{\hbar^{2}\mathbf{p}^{2}},
\end{equation}
which replaces the bare interaction strength $g$ with the $s$-wave
scattering length $a$. Here, $m_{r}\equiv mm_{I}/(m+m_{I})$ is the
reduced mass.

In the single-impurity limit, the model Hamiltonian can be conveniently
solved by the non-self-consistent $T$-matrix theory \cite{Combescot2007},
where the motion of the impurity is described by ladder diagrams,
accounting for the successive forward scatterings between the impurity
and fermions in the particle-particle channel. This gives rise to
the inverse two-particle vertex function at nonzero temperature $T$,
\begin{equation}
\Gamma^{-1}\left(\mathcal{Q}\right)=\frac{1}{g}+\sum_{\mathcal{K}}\mathcal{G}\left(\mathcal{K}\right)G_{0}\left(\mathcal{Q}-\mathcal{K}\right),\label{eq:vertexfunction0}
\end{equation}
where $\mathcal{K}\equiv(\mathbf{k},i\omega_{m})$ and $\mathcal{Q}\equiv(\mathbf{q},i\nu_{n})$
are the short-hand notations for momentum ($\mathbf{k}$ or $\mathbf{q}$)
and Matsubara frequency ($\omega_{m}=(2m+1)\pi k_{B}T$ or $\nu_{n}=2n\pi k_{B}T$
with integers $m$ and $n$), $\sum_{\mathcal{K}}\equiv k_{B}T\sum_{i\omega_{m}}\sum_{\mathbf{k}}$
and $\sum_{\mathcal{Q}}\equiv k_{B}T\sum_{i\nu_{n}}\sum_{\mathbf{q}}$,
and 
\begin{equation}
\mathcal{G}\mathcal{\left(K\right)}=\frac{1}{i\omega_{m}-\epsilon_{\mathbf{k}}+\mu}=\frac{1}{i\omega_{m}-\xi_{\mathbf{k}}}
\end{equation}
and
\begin{equation}
G_{0}\mathcal{\left(Q-K\right)}=\frac{1}{i\nu_{n}-i\omega_{m}-\epsilon_{\mathbf{q}-\mathbf{k}}^{(I)}}
\end{equation}
are the finite-temperature Green functions for fermionic atoms and
the impurity, respectively. 

In the single-impurity case, the Green function of atoms is barely
affected, so it takes the standard non-interacting form with a (temperature-dependent)
chemical potential $\mu(T)$. Instead, the impurity Green function
will be strongly renormalized by the impurity-atom coupling. However,
in the non-self-consistent $T$-matrix approach we take its non-interacting
form. This is because, there is a cancellation between the self-energy
renormalization of the impurity Green function and the vertex correction
to $\Gamma(\mathcal{Q})$ \cite{Mahan1967,Roulet1969,Nozieres1969}.
If we wish to use the impurity Green function in a self-consistent
way, we may then need to simultaneously take into account the vertex
correction, which is beyond the scope of this work. Otherwise, the
approximate theory may become worse. This explains why at zero temperature
we can obtain a very accurate (attractive) polaron energy within the
non-self-consistent $T$-matrix theory (or equivalently within Chevy's
variational approach) \cite{Combescot2007}. We note also that, with
a single impurity, the quantum statistics of the impurity is irrelevant.
Here, for concreteness we consider fermionic impurities, in connection
with the recent experiments \cite{Scazza2017,Zan2019}. Therefore,
in the two-particle vertex function $\Gamma(\mathcal{Q})$ we use
bosonic Matsubara frequencies $\nu_{n}=2n\pi k_{B}T$ ($n\in\mathbb{Z}$).

Given the free Green functions $\mathcal{G}(\mathcal{K})$ and $G_{0}(\mathcal{Q}-\mathcal{K})$
the Matsubara frequency summation in the vertex function $\Gamma(\mathcal{Q})$
over $i\omega_{m}$ is easy to evaluate. By replacing further the
bare interaction strength $g$ with the scattering length $a$, we
obtain \cite{Combescot2007},
\begin{equation}
\Gamma^{-1}\left(\mathcal{Q}\right)=\frac{m_{r}}{2\pi\hbar^{2}a}-\sum_{\mathbf{k}}\left[\frac{1-f\left(\xi_{\mathbf{k}}\right)}{i\nu_{n}-\xi_{\mathbf{k}}-\epsilon_{\mathbf{q}-\mathbf{k}}^{(I)}}+\frac{2m_{r}}{\hbar^{2}\mathbf{k}^{2}}\right],\label{eq:vertexfunction}
\end{equation}
where $f(x)\equiv1/(e^{\beta x}+1)$ with $\beta\equiv1/(k_{B}T)$
is the Fermi-Dirac distribution function. Here, we have discarded
the distribution function related to the impurity, as the impurity
chemical potential (which we do not show explicitly for convenience)
tends to $-\infty$ at nonzero temperature. This treatment is also
consistent with the fact that the quantum statistics of the impurity
is not relevant.

We can now calculate the self-energy of the impurity $\Sigma(\mathcal{K})$
by winding back the out-going leg of fermionic atoms in the vertex
function $\Gamma(\mathcal{Q})$ and connecting it with the in-coming
leg for fermionic atoms. This physically describes the hole excitation
and gives rise to, 
\begin{equation}
\Sigma\mathcal{\left(K\right)}=\sum_{\mathcal{Q}}\Gamma\left(\mathcal{Q}\right)\frac{1}{i\nu_{n}-i\omega_{m}-\xi_{\mathbf{q}-\mathbf{k}}}.
\end{equation}
To proceed, we can use the formal spectral representation of the vertex
function, i.e., 
\begin{equation}
\Gamma\left(\mathcal{Q}\right)=\intop_{-\infty}^{\infty}\frac{d\Omega}{\pi}\frac{\left[-\textrm{Im}\Gamma_{R}\left(\mathbf{q},\Omega\right)\right]}{i\nu_{n}-\Omega},
\end{equation}
where $\Gamma_{R}(\mathbf{q},\Omega)=\Gamma(\mathbf{q},i\nu_{n}\rightarrow\Omega+i0^{+})$
is the retarded vertex function after analytic continuation. The Matsubara
frequency summation in the self-energy over $i\nu_{n}$ again is straightforward
to evaluate. We find that,
\begin{equation}
\Sigma\mathcal{\left(K\right)}=\sum_{\mathbf{q}}\intop_{-\infty}^{\infty}\frac{d\Omega}{\pi}\left[-\textrm{Im}\Gamma_{R}\left(\mathbf{q},\Omega\right)\right]\frac{f\left(\xi_{\mathbf{q}-\mathbf{k}}\right)+f_{B}\left(\Omega\right)}{i\omega_{m}+\xi_{\mathbf{q}-\mathbf{k}}-\Omega},\label{eq:selfenergyFiniteDensity}
\end{equation}
where $f_{B}(\Omega)=1/(e^{\beta\Omega}-1)$ is the Bose-Einstein
distribution function. In the single-impurity limit, the molecule
occupation $f_{B}(\Omega)$ scales like $1/V$ and therefore is vanishingly
small, in line with the infinitely negative impurity chemical potential
as mentioned earlier. By removing $f_{B}\left(\Omega\right)$ in the
above expression, we again use the spectral representation and finally
arrive at \cite{Combescot2007},
\begin{equation}
\Sigma\mathcal{\left(K\right)}=\sum_{\mathbf{q}}f\left(\xi_{\mathbf{q}-\mathbf{k}}\right)\Gamma\left(\mathbf{q},i\omega_{m}+\xi_{\mathbf{q}-\mathbf{k}}\right).\label{eq:selfenergy}
\end{equation}

The retarded interacting impurity Green function then takes the form,
\begin{equation}
G_{R}(\mathbf{k},\omega)=\frac{1}{\omega-\epsilon_{\mathbf{k}}^{(I)}-\Sigma_{R}(\mathbf{k},\omega)},\label{eq:impurityGF}
\end{equation}
where the retarded self-energy $\Sigma_{R}(\mathbf{k},\omega)\equiv\Sigma(\mathbf{k},i\omega_{m}\rightarrow\omega+i0^{+})$.
The pole position of the impurity Green function determines the (attractive
or repulsive) polaron energy, i.e., 
\begin{equation}
\mathcal{E}_{P}\left(\mathbf{k}\right)=\epsilon_{\mathbf{k}}^{(I)}+\textrm{Re}\Sigma_{R}\left[\mathbf{k},\mathcal{E}_{P}\left(\mathbf{k}\right)\right].\label{eq:PolaronEnergy}
\end{equation}
By expanding the retarded self-energy near the zero momentum $\mathbf{k}=0$
and the polaron energy $\mathcal{E}_{P}\equiv\mathcal{E}_{P}\left(\mathbf{0}\right)$,
we calculate directly various quasiparticle properties, including
the polaron residue $\mathcal{Z}^{-1}=1-\partial\textrm{Re}\Sigma_{R}(0,\omega)/\partial\omega$,
the effective mass $m_{*}/m_{I}=\mathcal{Z}^{-1}/[1+\partial\textrm{Re}\Sigma_{R}(\mathbf{k},\mathcal{E}_{P})/\partial\epsilon_{\mathbf{k}}^{(I)}]$,
and also the polaron decay rate,
\begin{equation}
\Gamma=-2\mathcal{Z}\textrm{Im}\Sigma_{R}\left(\mathbf{0},\mathcal{E}_{P}\right).
\end{equation}
For a nonzero decay rate, it corresponds to the full width at half
maximum (FWHM) of the spectral function $A(\mathbf{k},\omega)\equiv-(1/\pi)\textrm{Im}G_{R}(\mathbf{k},\omega)$.

The set of equations, Eqs. (\ref{eq:vertexfunction}), (\ref{eq:selfenergy})
and (\ref{eq:impurityGF}), constitute the well-documented non-self-consistent
many-body $T$-matrix theory of Fermi polarons \cite{Massignan2014,Scazza2017,Combescot2007}.
At zero temperature, its equivalence to Chevy's variational ansatz
is well-known from the seminal work \cite{Combescot2007}. At finite
temperature, the equivalence to the finite-temperature variational
approach proposed by Meera Parish and her co-workers has also been
discussed \cite{Liu2019}.

\subsection{Numerical calculations}

The numerical solution of those coupled equations, however, is non-trivial,
particularly at finite temperature. The calculation of the retarded
self-energy at large frequency $\omega$ is subtle, due to the existence
of the two-particle continuum. As we can see from Eq. (\ref{eq:vertexfunction}),
the integrand has infinite number of poles once $i\nu_{n}\rightarrow\Omega>\omega_{\textrm{th}}(\mathbf{q})=\min_{\mathbf{k}}\{\xi_{\mathbf{k}}+\epsilon_{\mathbf{q}-\mathbf{k}}^{(I)}\}$,
where $\omega_{\textrm{th}}(\mathbf{q})$ is the threshold to enter
the two-particle continuum at the momentum $\mathbf{q}$. We therefore
need to take Cauchy principle value of the integral. This numerical
difficulty does not arise in the finite-temperature variational approach
\cite{Liu2019,Liu2020,Parish2021}, where one solves for the discretized
variational wave-functions or parameters in momentum space. The discretization
however would require a small broadening factor, in order to recover
a continuous spectral function. 

Alternatively, we can solve the coupled equations with Matsubara frequencies.
This strategy has been used in our previous work at low temperature
\cite{Hu2018}, where we can expand the impurity Green function, in
terms of the small Matsubara frequency $\omega_{m}=(2m+1)k_{B}T$,
in order to calculate the quasiparticle properties. At high temperature
or in the calculation of the spectral function $A(\mathbf{k},\omega)$,
one needs to numerically take the analytic continuation \cite{Tajima2018,Tajima2019}.
Unfortunately, this procedure is not well-defined and may lead to
uncontrollable uncertainties in the spectral function \cite{Goulko2016,Haussmann2009}.

In this work, we solve the coupled equations for the retarded impurity
self-energy and Green function with real frequency, following the
same idea in Ref. \cite{Mulkerin2019}, where the non-self-consistent
$T$-matrix theory has been used to understand a Fermi polaron system
at finite impurity density. The finite density makes the numerical
calculations very time-consuming. For example, we have to keep the
bosonic distribution function $f_{B}\left(\Omega\right)$ in Eq. (\ref{eq:selfenergyFiniteDensity})
and hence have an additional integral over the frequency $\Omega$.
As a result, one can hardly explore the finite temperature properties
of Fermi polarons. Here, we take the advantage of the single-impurity
limit and thereby greatly shorten the time needed for an accurate
calculation of the spectral function. There is also no need to introduce
any small broadening factor in the spectral function, allowing us
to make a comparison of our theoretical predictions with the recent
experiments \cite{Scazza2017,Zan2019}, without adjustable free parameters,
as we shall see.

\begin{figure*}
\begin{centering}
\includegraphics[clip,width=1\textwidth]{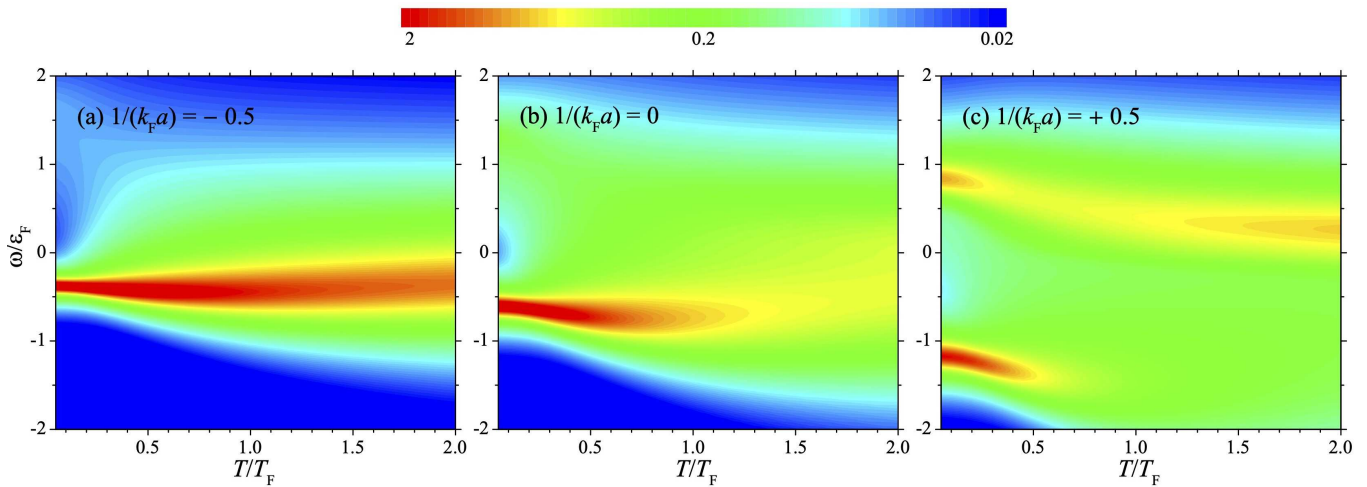}
\par\end{centering}
\caption{\label{fig1_akw2dttf} Two-dimensional contour plots of the temperature
evolution of the zero-momentum impurity spectral function $A(\mathbf{k}=0,\omega$)
at three dimensionless coupling strengths: $1/(k_{F}a)=-0.5$ (a),
$0$ (b), and $+0.5$ (c). The spectral function is in units of $\varepsilon_{F}^{-1}$,
where $\varepsilon_{F}\equiv\hbar^{2}k_{F}^{2}/(2m)$ is the Fermi
energy. Each plot is shown in the logarithmic scale. We typically
take the equal mass for the impurity and atoms, $m_{I}=m$, unless
otherwise specified.}
\end{figure*}

To be specific, we take the Fermi wave-vector $k_{F}\equiv(6\pi^{2}n)^{1/3}$
and Fermi energy $\varepsilon_{F}\equiv\hbar^{2}k_{F}^{2}/(2m)$ as
the units of the momentum (or wave-vector) and energy, respectively.
The temperature is then measured in units of the Fermi temperature
$T_{F}=\varepsilon_{F}/k_{B}$. This choice corresponds to set $2m=\hbar=k_{B}=1$.
We also define a mass ratio $-1<\alpha=(m_{I}-m)/(m_{I}+m)<1$, so
$2m_{I}=(1+\alpha)/(1-\alpha)$ and the reduced mass $2m_{r}=(1+\alpha)/2$.
The coupling between fermionic atoms and the impurity is characterized
by a dimensionless interaction parameter $1/(k_{F}a)$. We then find
that the dimensionless retarded two-particle propagator $\chi_{R}(\mathbf{q},\Omega)=\Gamma_{R}^{-1}(\mathbf{q},\Omega)$
can be written into the two-body and many-body parts (i.e., $\chi_{R}=\chi_{R}^{(2b)}+\chi_{R}^{(mb)}$),\begin{widetext}
\begin{equation}
\chi_{R}^{(2b)}\left(\mathbf{q},\Omega\right)=\frac{m_{r}}{2\pi\hbar^{2}a}-\sum_{\mathbf{k}}\left[\frac{1}{\Omega^{+}-\xi_{\mathbf{k}}-\epsilon_{\mathbf{q}-\mathbf{k}}^{(I)}}+\frac{2m_{r}}{\hbar^{2}\mathbf{k}^{2}}\right]=\frac{1+\alpha}{8\pi k_{F}a}+\frac{i\left(1+\alpha\right)^{3/2}}{8\sqrt{2}\pi}\sqrt{\Omega+\mu-\frac{\left(1-\alpha\right)}{2}q^{2}}\label{eq:kappa2b}
\end{equation}
and
\begin{equation}
\chi_{R}^{(mb)}\left(\mathbf{q},\Omega\right)=\sum_{\mathbf{k}}\frac{f\left(\xi_{\mathbf{k}}\right)}{\Omega^{+}-\xi_{\mathbf{k}}-\epsilon_{\mathbf{q}-\mathbf{k}}^{(I)}}=\intop_{0}^{\infty}\frac{k^{2}dk}{2\pi^{2}}\intop_{-1}^{+1}\frac{dx}{2}\frac{f\left(k^{2}-\mu\right)}{\left(\Omega^{+}+\mu\right)-k^{2}-\left[\left(1-\alpha\right)/\left(1+\alpha\right)\right]\left(q^{2}+k^{2}-2qkx\right)},\label{eq:kappamb}
\end{equation}
\end{widetext}where $\Omega^{+}\equiv\Omega+i0^{+}$ and $x\equiv\cos(\theta_{\mathbf{qk}})$,
with $\theta_{\mathbf{qk}}$ being the angle between the vectors $\mathbf{q}$
and $\mathbf{k}$. In Appendix A, we list the detailed expressions
for $\textrm{Im}\chi_{R}^{(mb)}$ and $\textrm{Re}\chi_{R}^{(mb)}$
and discuss their qualitative feature. Once the two-particle propagator
$\chi_{R}(\mathbf{q},\Omega)$ is calculated, we take the inverse
to obtain the real and imaginary parts of the two-particle vertex
function $\Gamma_{R}(\mathbf{q},\Omega)$, which physically serves
as the Green function of molecules. It is then straightforward to
calculate the retarded self-energy,
\begin{equation}
\Sigma_{R}(\mathbf{k},\omega)=\intop_{0}^{\infty}\frac{q^{2}dq}{2\pi^{2}}\intop_{-1}^{+1}\frac{dx}{2}f\left(\xi_{\mathbf{q}-\mathbf{k}}\right)\Gamma_{R}\left(\mathbf{q},\omega+\xi_{\mathbf{q}-\mathbf{k}}\right),
\end{equation}
where $\xi_{\mathbf{q}-\mathbf{k}}=q^{2}+k^{2}-2qkx-\mu$. This two-dimensional
integral can be efficiently determined, if we take care of the possible
pole in the two-particle vertex function, which corresponds to the
two-body bound state that may arise in the strong-coupling regime
when the scattering length is positive, $a>0$.

\section{Spectral function and quasiparticle lifetime}

In this section, we discuss in detail the finite-temperature spectral
function of Fermi polarons near a Feshbach resonance, in the case
of equal mass $m_{I}=m$ (i.e., $\alpha=0$). We also present the
quasiparticle properties of both attractive and repulsive polarons
at arbitrary temperatures, extending our previous low-temperature
results on attractive polarons \cite{Hu2018}. We emphasize that we
do not use any small broadening factor in numerical calculations,
so the quasiparticle decay rate obtained determines the intrinsic
lifetime of Fermi polarons.

\begin{figure}
\begin{centering}
\includegraphics[width=0.45\textwidth]{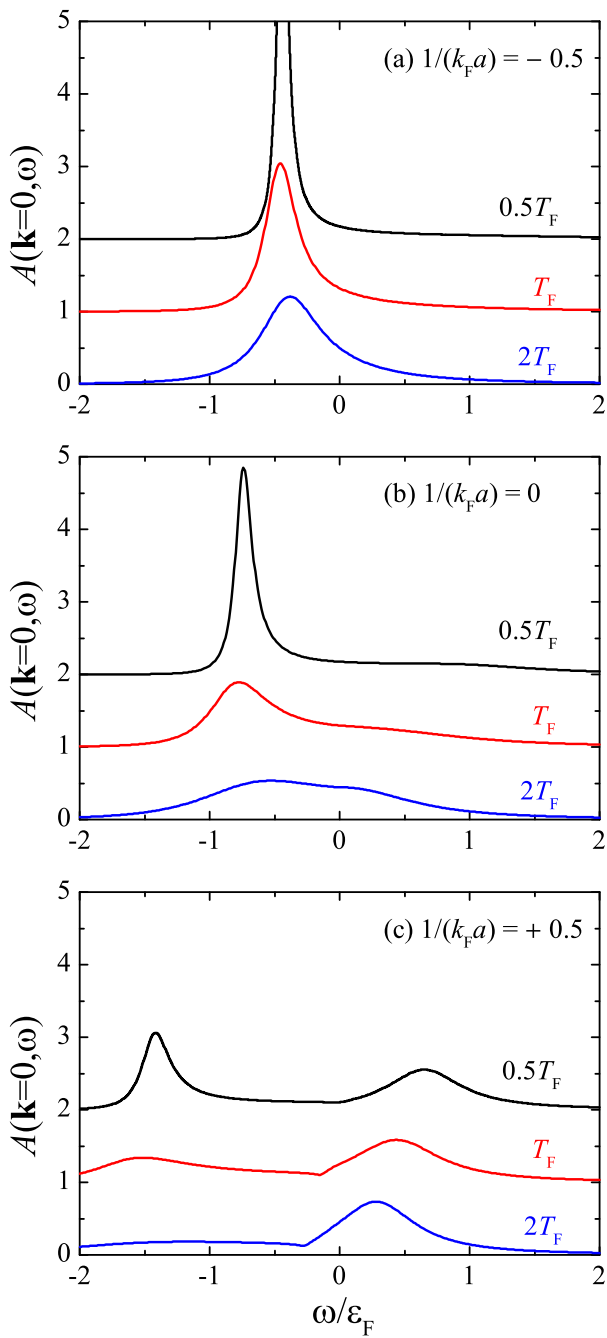}
\par\end{centering}
\caption{\label{fig2_akwttf} The impurity spectral function $A(\mathbf{k}=0,\omega$)
at the dimensionless coupling strengths: $1/(k_{F}a)=-0.5$ (a), $0$
(b), and $+0.5$ (c). These curves are the cuts at three different
temperatures (i.e., $T/T_{F}=0.5$, $1$, and $2$) in Fig. \ref{fig1_akw2dttf}. }
\end{figure}

Fig. \ref{fig1_akw2dttf} presents the two-dimensional contour plots
of the zero-momentum spectral function $A(\mathbf{k}=0,\omega$) as
a function of temperature at the crossover from a Bardeen--Cooper--Schrieffer
(BCS) superfluid to a Bose-Einstein condensate (BEC), where the dimensionless
interaction parameter changes from the BCS side (a, $1/(k_{F}a)=-0.5$),
to the unitary limit (b, $1/(k_{F}a)=0$), and finally to the BEC
side (b, $1/(k_{F}a)=+0.5$). A few example traces at fixed temperatures
through the three contour plots are shown in Fig. \ref{fig2_akwttf}.
These traces are often referred to as energy distribution curves,
or EDCs.

On the BCS side, we find a well-defined attractive Fermi polaron at
all the temperatures considered in this work (i.e., up to $2T_{F}$).
At absolutely zero temperature, it is well-known the ground-state
Fermi polaron exhibits itself as a delta-function peak in the spectral
function. A nonzero temperature typically brings a thermal broadening
to the quasiparticle peak. At $1/(k_{F}a)=-0.5$, the thermal broadening
turns out to have a weak temperature dependence. As shown in Fig.
\ref{fig2_akwttf}(a), at the Fermi degenerate temperature $T_{F}$
the attractive polaron remains as a reasonably sharp peak in the spectrum.
Even at the largest temperature $2T_{F}$, we still find a well-preserved
Lorentzian shape with a width at about half Fermi energy only. The
quasiparticle peak position also depends very weakly on the temperature.
We only notice a very slight shift in the peak position above Fermi
degeneracy $T>T_{F}$.

In the unitary limit, the situation dramatically changes. At very
low temperatures, we can see clearly from Fig. \ref{fig1_akw2dttf}(b)
an incoherent broad distribution well above the ground-state attractive
polaron peak at $\mathcal{E}_{P}\simeq-0.6\varepsilon_{F}$. In between,
there is an area with very low spectral weight, which is named as
dark continuum in the literature \cite{Goulko2016}. The broad distribution
might be viewed as a precursor of repulsive polarons. By increasing
temperature, the dark continuum gradually disappears. At the same
time, the attractive polaron peak shifts downwards and becomes broader.
At $T=T_{F}$, the width of the attractive polaron peak is comparable
to the Fermi energy and the peak is setting on an incoherent background
(see the red curve in Fig. \ref{fig2_akwttf}(b)). By further increasing
temperature, the attractive polaron essentially dissolves. At $2T_{F}$,
we find the remnant of the attractive polaron merges with an enhanced
incoherent background. Both of them seem to distribute symmetrically
around zero energy (see the blue curve in Fig. \ref{fig2_akwttf}(c)).

On the BEC side with $1/(k_{F}a)=0.5$, at low temperatures the precursor
of repulsive polaron develops into a well-defined quasiparticle at
the energy $\mathcal{E}_{P}\simeq0.8\varepsilon_{F}$, as we can see
from Fig. \ref{fig1_akw2dttf}(c). Both attractive and repulsive polarons
have a red-shift in their energy, with increasing temperature. We
find that the attractive polaron quickly disappears at temperature
around $0.7T_{F}$. In sharp contrast, the repulsive polaron remains
very robust with temperature. Apart form a systematic downshift of
the peak position, its profile remains essentially the same, as seen
from Fig. \ref{fig2_akwttf}(c). There is also a slight reduction
in the width of the repulsive polaron with increasing temperature,
which again indicates the robustness of the repulsive polaron against
thermal fluctuations.

\begin{figure}
\begin{centering}
\includegraphics[clip,width=0.48\textwidth]{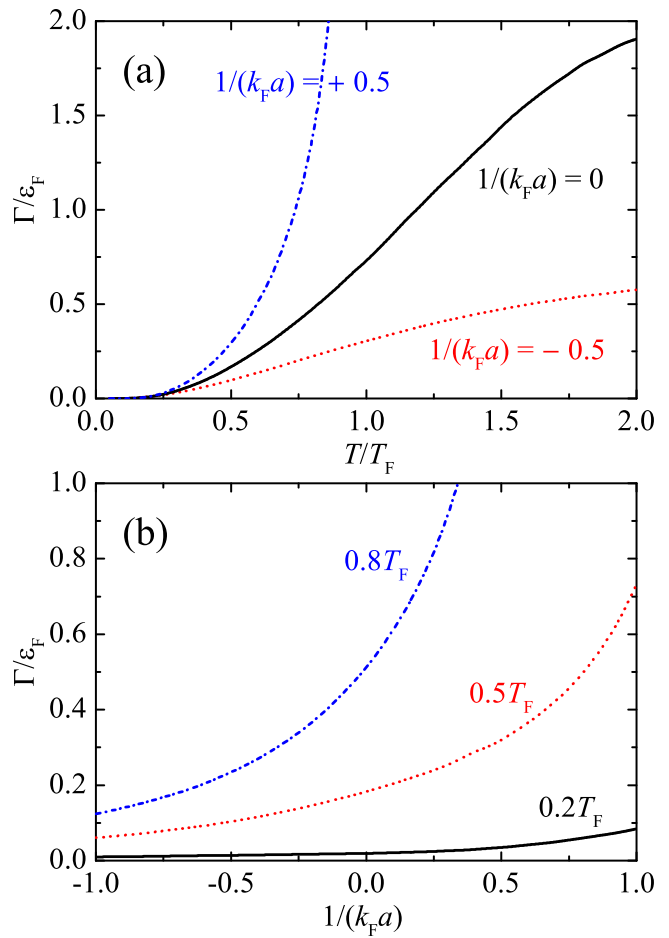}
\par\end{centering}
\caption{\label{fig3_AttractiveDecayRT} Quasiparticle lifetime of attractive
polarons as a function of temperature at fixed coupling strengths
(a) and as a function of coupling strength at given temperatures (b).}
\end{figure}

\begin{figure}[t]
\begin{centering}
\includegraphics[width=0.48\textwidth]{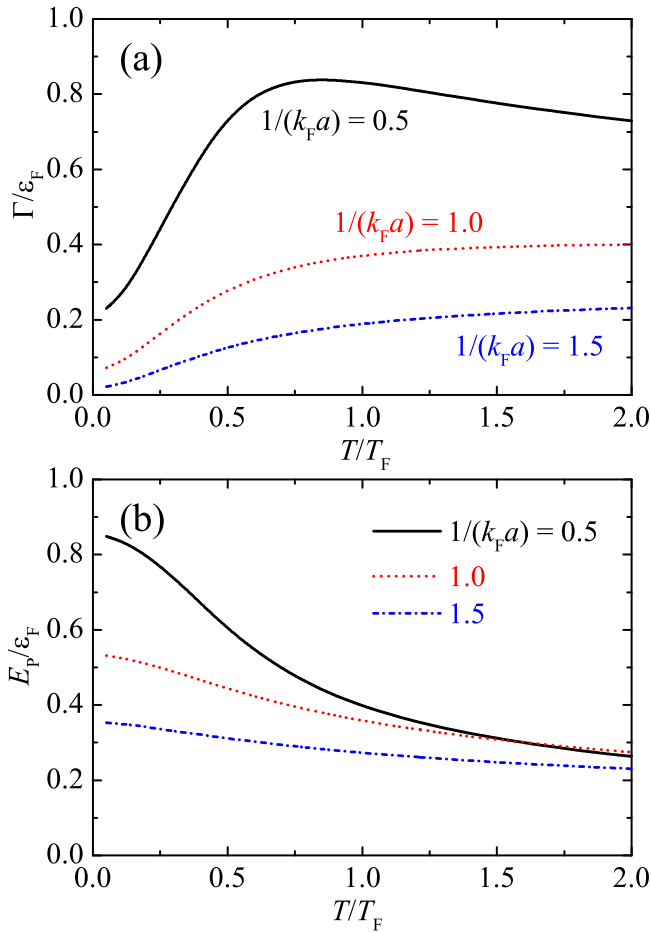}
\par\end{centering}
\caption{\label{fig4_RepulsiveDecayRT} Quasiparticle lifetime (a) and energy
(b) of repulsive polarons as a function of temperature at fixed coupling
strengths: $1/(k_{F}a)=0.5$ (black solid lines), $1.0$ (red dotted
lines), and $1.5$ (blue dot-dashed lines).}
\end{figure}

In Fig. \ref{fig3_AttractiveDecayRT}, we show the decay rate $\Gamma$
of attractive polarons, which corresponds to the FWHM of the attractive
polaron peak if it exists, as functions of temperature and interaction
parameter. In general, the decay rate increases with increasing both
$T$ (Fig. \ref{fig3_AttractiveDecayRT}(a)) and $1/(k_{F}a)$ (Fig.
\ref{fig3_AttractiveDecayRT}(b)). At low temperatures, the decay
rate has a $T^{2}$-dependence, according to the Fermi liquid theory
\cite{FermiLiquidTheory}. In the unitary limit, we find numerically
that $\Gamma/\varepsilon_{F}\simeq0.75(T/T_{F})^{2}$ for $T<T_{F}$.
Above the Fermi degeneracy, there is a clear deviation from the $T^{2}$
law, indicating the breakdown of the quasiparticle description in
terms of Fermi polarons.

In Fig. \ref{fig4_RepulsiveDecayRT}(a), we report the decay rate
of repulsive polarons on the BEC side. The decay rate typically decreases
with increasing interaction parameter $1/(k_{F}a)$. It also generally
increases with increasing temperature. An exception occurs close to
the Feshbach resonance. For $1/(k_{F}a)=0.5$, we find that there
is non-monotonic dependence of the decay rate on the temperature.
The decay rate attains a maximum value $\sim0.8\varepsilon_{F}$ at
about $0.8T_{F}$. Above this temperature, the decay rate starts to
slowly decrease. The slow decrease is in line with the observation
of a slight reduction in the spectral width of the repulsive polaron
peak in Fig. \ref{fig2_akwttf}(c) at large temperature. In Fig. \ref{fig4_RepulsiveDecayRT}(b),
we also report the energy of repulsive polarons, which decreases monotonically
with increasing temperature.

\begin{figure}[t]
\begin{centering}
\includegraphics[width=0.5\textwidth]{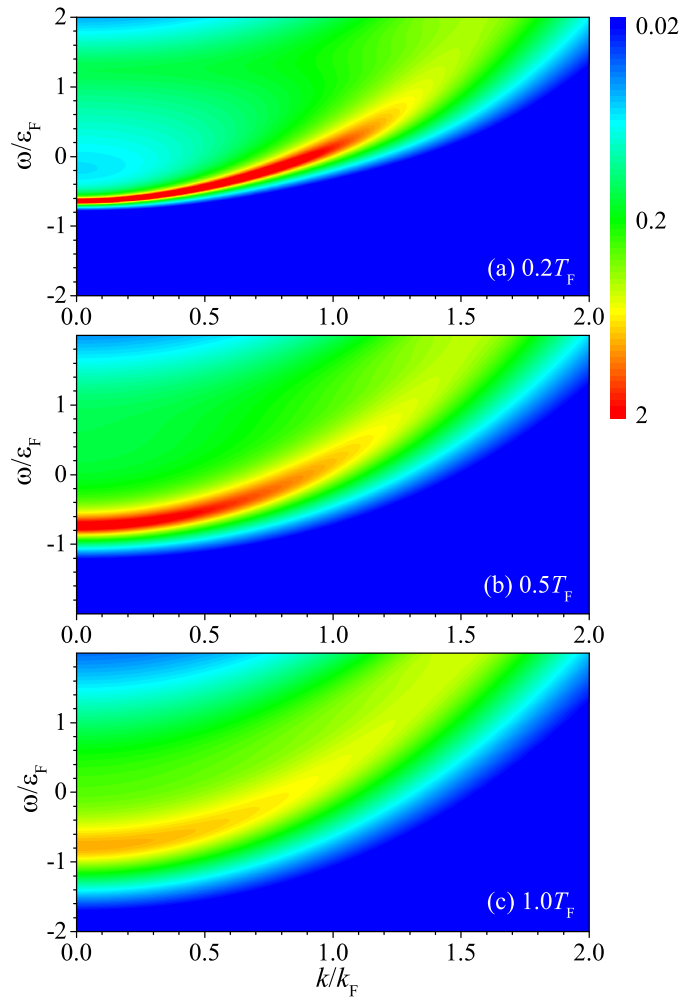}
\par\end{centering}
\caption{\label{fig5_akw2dkk} The two-dimensional contour plot of the finite-momentum
impurity spectral function $A(\mathbf{k},\omega$) at three temperatures:
$T/T_{F}=0.2$ (a), $0.5$ (b), and $1.0$ (c). Each plot is shown
in the logarithmic scale. We consider the unitary coupling strength
$1/(k_{F}a)=0$.}
\end{figure}

As a brief conclusion of this section, in the weak-coupling regime
(i.e., the BCS side for attractive polarons and the BEC side for repulsive
polarons), the polaron quasiparticle is robust against thermal fluctuations
and remains well-defined above Fermi degeneracy. While in the strong-coupling
regime or the unitary limit, attractive Fermi polaron exists at $T\lesssim T_{F}$.
Above this characteristic temperature, we find a co-existence of the
remnant of the attractive Fermi polaron and of an enhanced incoherent
background. The latter might be understood as the precursor of a repulsive
polaron at large temperature.

\section{Ejection rf-spectroscopy in the unitary limit}

In experiments, quasiparticle properties of Fermi polarons can be
conveniently measured by using either ejection or injection rf-spectroscopy
\cite{Schirotzek2009,Koschorreck2012,Kohstall2012,Scazza2017,Zan2019}.
In most cases, the measured spectroscopy is not momentum resolved,
since the density of impurities should be dilute enough, which sets
a limitation on signal that brings difficulty for resolving the momentum
distribution of transferred atoms. In other words, the spectroscopy
is an averaged spectral function over all momenta. In Fig. \ref{fig5_akw2dkk},
we show the spectral function of a unitary Fermi polaron in the $\omega$-$k$
plane at three typical temperatures. The non-trivial momentum dependence
of the spectral function suggests that we may need to carefully examine
the dependence of the rf-spectroscopy on the density of impurities.

In more detail, in the ejection rf-spectrosocpy scheme, a system of
strongly interacting Fermi polarons is initially prepared and then
a rf pulse transfers impurities to a third, unoccupied hyperfine state.
In the absence of the final-state effect (i.e., the transferred impurity
atom does not interact with the Fermi sea) and in the linear response
regime, the transfer rate as a function of the energy $\omega$, defined
as the ejection rf spectrum, is given by \cite{Mulkerin2019,Liu2020,Torma2014,Punk2007,NoteImpuritymu},
\begin{equation}
I\left(\omega\right)=\sum_{\mathbf{k}}A\left[\mathbf{k},\epsilon_{\mathbf{k}}^{(I)}-\omega\right]f\left(\epsilon_{\mathbf{k}}^{(I)}-\omega-\mu_{I}\right).
\end{equation}
Here, $\mu_{I}$ is the impurity chemical potential. The introduction
of $\mu_{I}$ and the Fermi-Dirac distribution function is necessary
to account for the finite impurity density $n_{\textrm{imp}}$ in
experiments:
\begin{equation}
n_{\textrm{imp}}=\sum_{\mathbf{k}}\intop_{-\infty}^{\infty}d\omega A\left(\mathbf{k},\omega\right)f\left(\omega-\mu_{I}\right).
\end{equation}
It is easy to see that the ejection rf spectrum is normalized to the
impurity density, i.e., $\int d\omega I(\omega)=n_{\textrm{imp}}$.
In the single impurity limit, $n_{\textrm{imp}}\rightarrow0$ and
$\mu_{I}\rightarrow-\infty$ at finite temperature. In this idealized
case, we may replace the Fermi-Dirac distribution function by a classical
Boltzmann distribution \cite{Liu2020}, $f(\omega-\mu_{I})\simeq e^{-\beta\omega}e^{\beta\mu_{I}}$.
Therefore, we obtain,
\begin{eqnarray}
n_{\textrm{imp}} & = & e^{\beta\mu_{I}}\sum_{\mathbf{k}}\intop_{-\infty}^{\infty}d\omega e^{-\beta\omega}A\left(\mathbf{k},\omega\right),\\
I\left(\omega\right) & = & e^{\beta\mu_{I}}\sum_{\mathbf{k}}e^{-\beta\epsilon_{\mathbf{k}}^{(I)}}e^{\beta\omega}A\left[\mathbf{k},\epsilon_{\mathbf{k}}^{(I)}-\omega\right].
\end{eqnarray}
By removing the unknown impurity fugacity $e^{\beta\mu_{I}}$, we
arrive at an elegant expression first derived by Meera Parish and
her co-workers \cite{Liu2020},
\begin{equation}
\frac{I\left(\omega\right)}{n_{\textrm{imp}}}=e^{\beta\mathcal{F}}e^{\beta\omega}\sum_{\mathbf{k}}e^{-\beta\epsilon_{\mathbf{k}}^{(I)}}A\left[\mathbf{k},\epsilon_{\mathbf{k}}^{(I)}-\omega\right],\label{eq:ejectionRF}
\end{equation}
where the quantity $\mathcal{F}(T)$ define by \cite{NoteFreeEnergy}
\begin{equation}
e^{-\beta\mathcal{F}}\equiv\sum_{\mathbf{k}}\intop_{-\infty}^{\infty}d\omega e^{-\beta\omega}A\left(\mathbf{k},\omega\right)
\end{equation}
can be physically interpreted as the impurity free energy. This is
readily seen in the free-particle limit, where $A(\mathbf{k},\omega)=\delta(\omega-\epsilon_{\mathbf{k}}^{(I)})$
and $\mathcal{F}_{0}(T)$ given by $e^{-\beta\mathcal{F}_{0}}\equiv\sum_{\mathbf{k}}e^{-\beta\epsilon_{\mathbf{k}}^{(I)}}$
is the free energy of a free particle.

\begin{figure}
\begin{centering}
\includegraphics[width=0.48\textwidth]{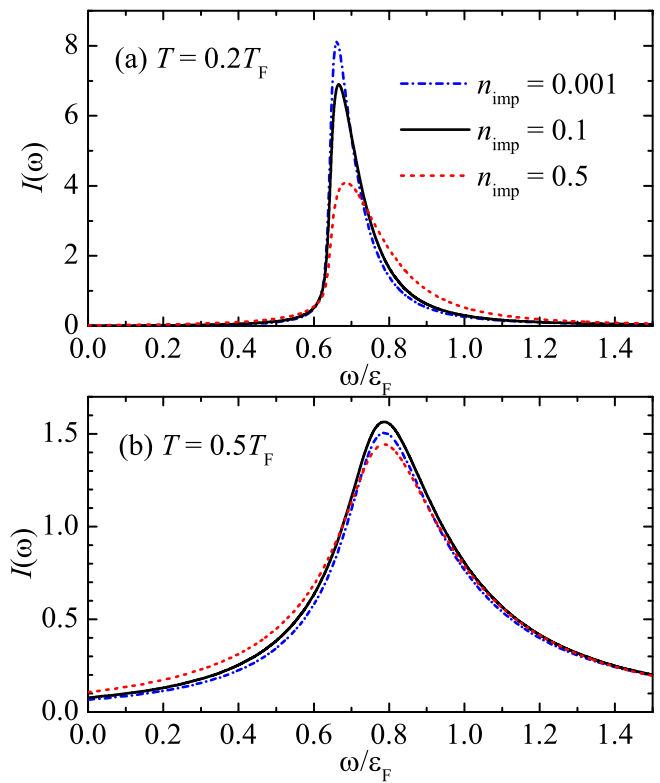}
\par\end{centering}
\caption{\label{fig6_RFnimp} The ejection rf spectra of a unitary Fermi polaron
at temperatures $T=0.2T_{F}$ (a) and $T=0.5T_{F}$ (b), at three
impurity densities as indicated. The curves are normalized to unity,
i.e., $\int d\omega I(\omega)=1$. This can be simply achieved by
dividing $I(\omega)$ the impurity density $n_{\textrm{imp}}$. }
\end{figure}

Let us focus on the unitary coupling. In Fig. \ref{fig6_RFnimp},
we examine the density dependence of the ejection rf spectra of Fermi
polarons in the unitary limit at two characteristic temperatures.
The curves with $n_{\textrm{imp}}/n=0.001$ can be well-understood
as the single-impurity limit. We find that the spectrum changes slightly
if we increase the impurity density to $n_{\textrm{imp}}/n=0.1$,
a typical density used in the recent MIT measurements \cite{Zan2019}.
This indicates that the polaron limit is well-reached in the experiment.
By further increasing impurity density to 0.5, the rf spectrum at
low temperature (a, $T=0.2T_{F}$) changes appreciably: the peak position
shifts to high energy and there is a significant broadening in the
line shape. At large temperature (b, $T=0.5T_{F}$), however, the
change is not notable. At this temperature, the effective reduced
temperature for impurity is about $T/[(0.5)^{2/3}T_{F}]\simeq0.8$,
which is close to the Fermi degeneracy. Therefore, the impurity may
already behave classically, following the Boltzmann distribution.
This explains the weak density dependence of the rf spectrum at high
temperature observed in Fig. \ref{fig6_RFnimp}(b).

\begin{figure}
\begin{centering}
\includegraphics[width=0.48\textwidth]{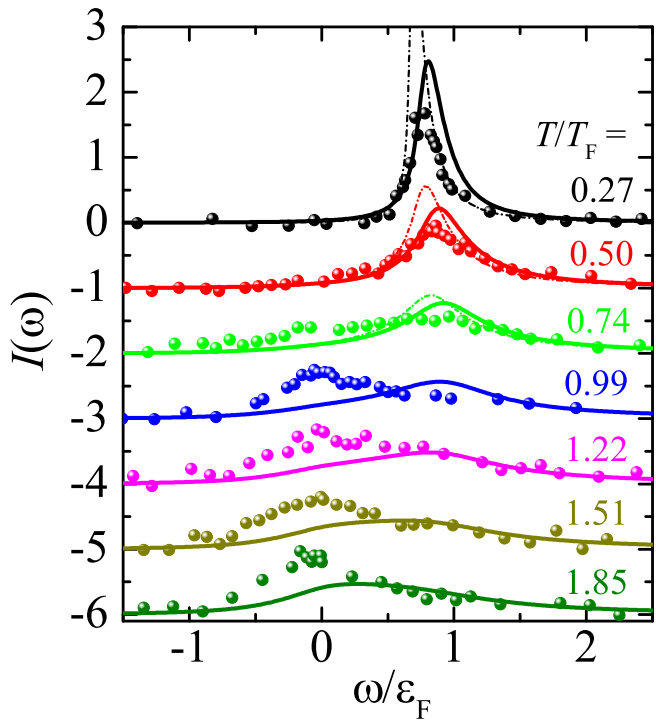}
\par\end{centering}
\caption{\label{fig7_RFexpt1} The comparison of the theory (lines) with the
experimental data from MIT (circles) \cite{Zan2019}, for the ejection
rf spectra of a unitary Fermi polaron at various temperatures (as
indicated). For clarity, the origin for the theoretical curves and
the experimental data at different temperatures has been down-shifted.
We have applied a Lorentzian broadening on all the solid theoretical
curves to take into account a well-calibrated experimental energy
resolution $0.1\varepsilon_{F}$ \cite{Zan2019}, and have also right-shifted
the curves by an amount $0.09\varepsilon_{F}$ to eliminate the residual
final-state effect \cite{Zan2019}. At the three lowest temperatures,
the theoretical curves without broadening and shift are shown by dot-dashed
lines. In the comparison, we do not include any adjustable free parameters.
The impurity density is taken as $n_{\textrm{imp}}=0.1$, following
the experimental condition \cite{Zan2019}. The spectra are normalized
to unity, $\int d\omega I(\omega)=1$.}
\end{figure}

In Fig. \ref{fig7_RFexpt1}, we compare our theoretical predictions
on the ejection rf spectrum with the data from the recent MIT experiment
\cite{Zan2019}, without any fitting parameters. To account for a
background energy resolution $0.1\varepsilon_{F}$ \cite{Zan2019},
we have taken a convolution of the theoretical curves with a Lorentzian
profile. There is also a weak final-state effect due to the residual
scattering length $a_{f}$ between the third hyperfine state and the
fermionic atom state, as characterized by an interaction parameter
$k_{F}a_{f}\simeq0.2$ \cite{Schirotzek2009,Zan2019}. In equilibrium,
the final state therefore is better described as a repulsive polaron
state with energy $\mathcal{E}_{P}^{(f)}\simeq[4k_{F}a_{f}/(3\pi)]\varepsilon_{F}\simeq0.09\varepsilon_{F}$
and a thermal (temperature-dependent) decay rate $\Gamma^{(f)}$ at
most a few percent of Fermi energy. By neglecting $\Gamma^{(f)}$,
this gives rise to a mean-field blue shift $0.09\varepsilon_{F}$
to the spectrum, which we have taken into account in the comparison.

\begin{figure}
\begin{centering}
\includegraphics[width=0.48\textwidth]{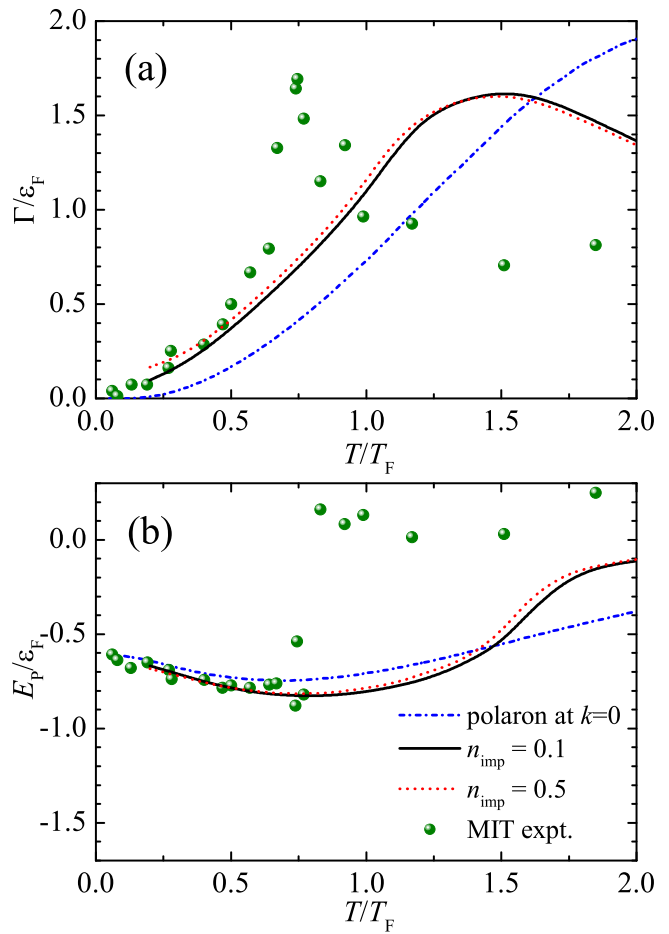}
\par\end{centering}
\caption{\label{fig8_RFexpt2} The FWHM width (a) and the peak position (b),
extracted from the rf spectra of a unitary Fermi polaron, as a function
of temperature. We compare the theoretical predictions (after spectral
broadening and shift) at $n_{\textrm{imp}}=0.1$ (black solid lines)
and $n_{\textrm{imp}}=0.5$ (red dotted lines), with the experimental
data from MIT (circles) \cite{Zan2019}. The blue dot-dashed lines
in (a) and (b) show the quasiparticle lifetime and energy of the ground-state
attractive polaron at zero momentum, respectively. }
\end{figure}

We find a good agreement between theory and experiment for the spectra
at positive energy (i.e., $\omega>0.5\varepsilon_{F}$), for \emph{all}
the temperatures considered in the comparison. As the positive energy
part of the ejection rf spectrum contributed mainly by attractive
polarons (or their remnant), this remarkable agreement clearly indicates
that our non-self-consistent $T$-matrix theory provides a satisfactory
description of ground-state attractive Fermi polarons at arbitrary
temperatures. Furthermore, at low temperatures (i.e., $T\leq0.5T_{F}$),
as the spectrum is dominated by the coherent quasiparticle contribution,
the agreement is good for all values of the energy.

At the temperature above $0.5T_{F}$, however, there is a growing
contribution in the experimental data, centered around the zero energy
$\omega\sim0$. Roughly speaking, this contribution might be understood
as the precursor of the excited branch of repulsive polarons, as we
discussed earlier. Our non-self-consistent theory seems to strongly
underestimate the magnitude of this excited branch. There are two
potential sources for the discrepancy: the lack of either the self-energy
renormalization or vertex renormalization in the theory. The former
is due to the use of a bare, non-interacting impurity Green function
in the vertex function (see Eq. (\ref{eq:vertexfunction0})). This
self-energy renormalization might be obtained by adopting a self-consistent
many-body $T$-matrix theory \cite{Hu2018}. The vertex renormalization
is more difficult to achieve, as we need to go beyond the ladder approximation
or the $T$-matrix framework \cite{Roulet1969}.

There are also two unlikely reasons for the discrepancy. The first
one is the final-state effect. Although we consider the mean-field
shift due to the scattering length $a_{f}$ (that corresponds to the
self-energy correction to the final state Green function), we also
need to examine the so-called Aslamazov-Larkin (AL) contribution to
the rf spectrum \cite{Haussmann2009,Pieri2009}. On the other hand,
as we consider the dilute polaron limit, we completely neglect the
residual repulsive interaction between polarons. Unfortunately, a
quantitative treatment of either the final-state effect or the polaron-polaron
interaction is difficult.

In Fig. \ref{fig8_RFexpt2}, we report the comparison for the width
and peak position, extracted from the experimental data \cite{Zan2019}
(circles) or from our theoretically predicted ejection rf spectra
(solid lines). Here, for clarity we have removed in the experimental
data the background energy resolution $0.1\varepsilon_{F}$ for the
width and the mean-field shift $0.09\varepsilon_{F}$ for the peak
position. The width is commonly understood as the decay rate of Fermi
polarons and the peak position in the ejection spectrum corresponds
to the polaron energy (i.e., $-\mathcal{E}_{P}$). This common interpretation
is only qualitatively useful, as the rf spectrum is not momentum-resolved.
To see this, we have included in Fig. \ref{fig8_RFexpt2} the decay
rate and polaron energy calculated at zero momentum (see the blue
dot-dashed lines). It is clear that the width extracted from the rf
spectrum differs significantly from the zero-momentum decay rate at
all temperature, due to the momentum average. The peak position seems
to agree with the zero-momentum polaron energy at very low temperatures.
However, there is an appreciable difference once the temperature $T>0.5T_{F}$. 

For the comparison between the theory and experiment for the width
and peak position, we find a quantitative agreement at $T\leq0.5T_{F}$,
consistent with the observation in Fig. \ref{fig7_RFexpt1}. Above
this temperature, the experimental width increases sharply and peaks
at about $0.75T_{F}$, at which the peak position suddenly jumps to
zero energy. All these features are related to the enhanced excited
branch of repulsive polarons, which unfortunately can not be account
for by our non-self-consistent many-body $T$-matirx theory, as we
emphasized earlier. 

It is worth noting that the theoretical width and peak position have
previously been calculated within similar non-self-consistent $T$-matrix
theory at finite impurity density \cite{Mulkerin2019,Tajima2019}.
An attempt \cite{Tajima2019} was also made to understand the MIT
data in Fig. \ref{fig8_RFexpt2}. Our results focus on the physical
limit of a single impurity, with the uncertainty from numerical analytic
continuation removed. Moreover, our new comparison for the ejection
rf spectra in Fig. \ref{fig7_RFexpt1} clearly reveals the key reason
for the discrepancy. That is, we need to find a more adequate theoretical
description for the excited branch of repulsive polarons at high temperature.

\begin{figure}
\begin{centering}
\includegraphics[width=0.48\textwidth]{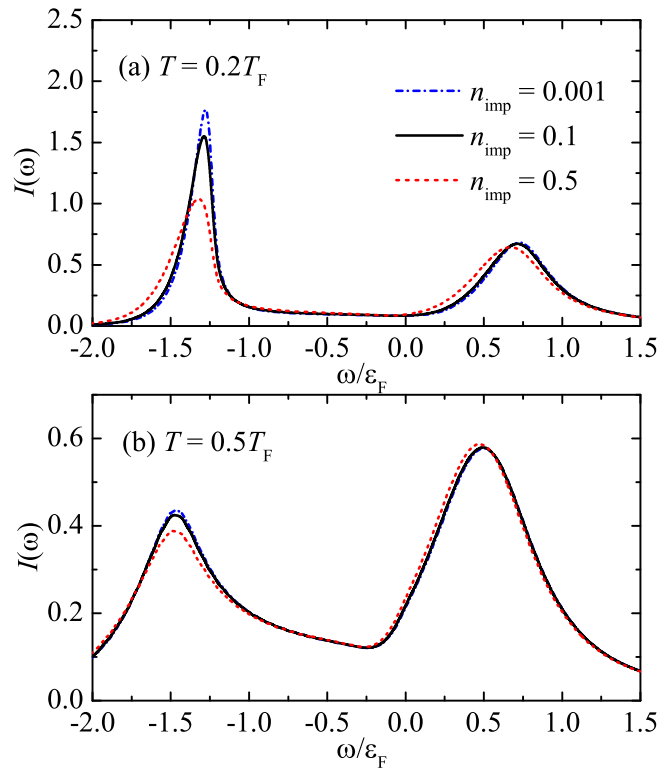}
\par\end{centering}
\caption{\label{fig9_injRFnimp1} The injection rf spectra of a Fermi polaron
at temperatures $T=0.2T_{F}$ (a) and $T=0.5T_{F}$ (b), at three
impurity densities as indicated. Here, we take the coupling strength
$1/(k_{F}a)=0.5$. The curves are normalized to the unity, $\int d\omega I(\omega)=1$.}
\end{figure}

\begin{figure}
\begin{centering}
\includegraphics[width=0.48\textwidth]{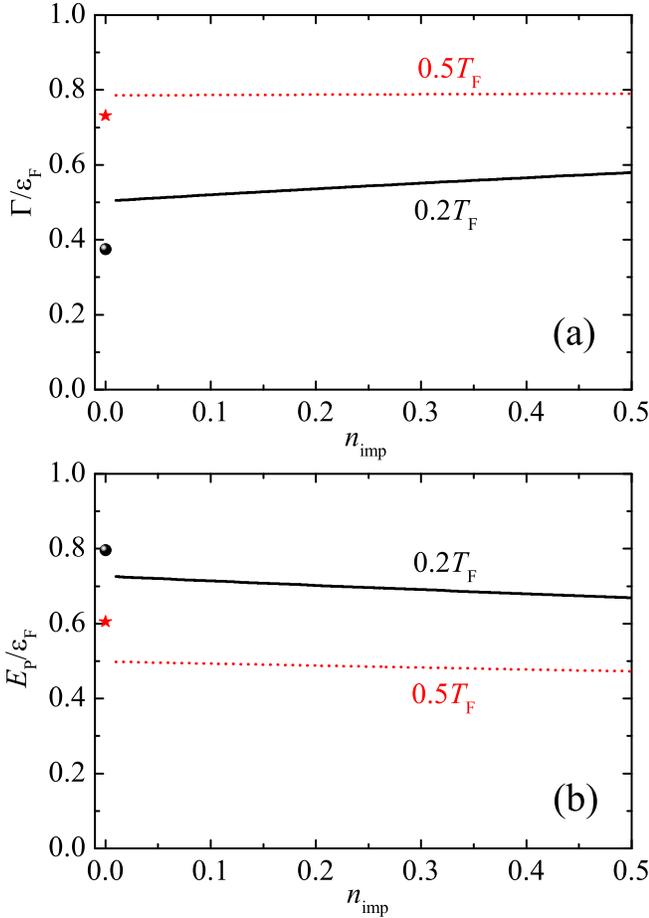}
\par\end{centering}
\caption{\label{fig10_injRFnimp2} The width (a) and the peak position (b),
extracted from the injection rf spectra at $1/(k_{F}a)=0.5$, as a
function of the impurity density $n_{\textrm{imp}}$ at temperatures
$T=0.2T_{F}$ (black solid lines) and $T=0.5T_{F}$ (red dotted lines).
The symbols at zero impurity density show the quasiparticle lifetime
or the energy of the zero-momentum repulsive polaron.}
\end{figure}

\section{Injection rf-spectroscopy of repulsive polarons}

Let us now turn to a reversed scheme of the injection rf-spectroscopy,
where the impurities initially occupy the non-interacting (or weakly-interacting)
third hyperfine state and are then transferred to the strongly-interacting
polaron state. This scheme is useful to probe the excited repulsive
polaron branch \cite{Scazza2017}, which can hardly be detected in
the standard ejection rf spectroscopy due to its negligible thermal
occupation. By neglecting the initial-state effect, the injection
rf spectrum is given by \cite{Mulkerin2019,Liu2020,Torma2014},
\begin{equation}
I\left(\omega\right)=\sum_{\mathbf{k}}A\left[\mathbf{k},\epsilon_{\mathbf{k}}^{(I)}+\omega\right]f\left(\epsilon_{\mathbf{k}}^{(I)}-\mu_{i}\right),
\end{equation}
where $\mu_{i}$ is the impurity chemical potential in the initial
third hyperfine state, to be determined by the number equation, $n_{\textrm{imp}}=\sum_{\mathbf{k}}f(\epsilon_{\mathbf{k}}^{(I)}-\mu_{i})$.
In the idealized single-impurity limit ($\mu_{i}\rightarrow-\infty$
at nonzero temperature), once again we can write $n_{\textrm{imp}}=e^{\beta\mu_{i}}\sum_{\mathbf{k}}e^{-\beta\epsilon_{\mathbf{k}}^{(I)}}$
and 
\begin{equation}
\frac{I\left(\omega\right)}{n_{\textrm{imp}}}=e^{\beta\mathcal{F}_{0}}\sum_{\mathbf{k}}e^{-\beta\epsilon_{\mathbf{k}}^{(I)}}A\left[\mathbf{k},\epsilon_{\mathbf{k}}^{(I)}+\omega\right].\label{eq:injectionRF}
\end{equation}
A comparison with Eq. (\ref{eq:ejectionRF}) gives us a very simple
relation between the ejection and injection rf spectra in the single-impurity
Boltzmann limit \cite{Liu2020},
\begin{equation}
I_{\textrm{ej}}\left(\omega\right)=e^{\beta\Delta\mathcal{F}}e^{\beta\omega}I_{\textrm{inj}}\left(-\omega\right),
\end{equation}
where $\Delta\mathcal{F}(T)\equiv\mathcal{F}-\mathcal{F}_{0}$ and
the subscripts ``ej'' and ``inj'' indicate the ejection and injection
spectra, respectively.

In Fig. \ref{fig9_injRFnimp1}, we show the injection rf spectra at
the interaction parameter $1/(k_{F}a)=0.5$ and at various impurity
densities. Two peaks are clearly visible at negative and positive
energies, contributed from the attractive and repulsive polarons,
respectively. For the attractive polaron peak, its density dependence
is similar to what we have seen in Fig. \ref{fig6_RFnimp}. For the
repulsive polaron peak, the density dependence turns out to be very
weak.

In Fig. \ref{fig10_injRFnimp2}, we examine in a more careful way
the weak density dependence of the width and peak position of repulsive
polarons. We are specifically interested in comparing the width and
peak position in the dilute limit with the decay rate and polaron
energy at zero momentum, which are indicated in the figure by symbols.
It is readily seen that, due to the momentum average in the injection
rf spectrum, in the single-impurity limit the width differs from the
zero-momentum decay rate and the peak position does not locate at
the polaron energy. 

\begin{figure}
\begin{centering}
\includegraphics[width=0.48\textwidth]{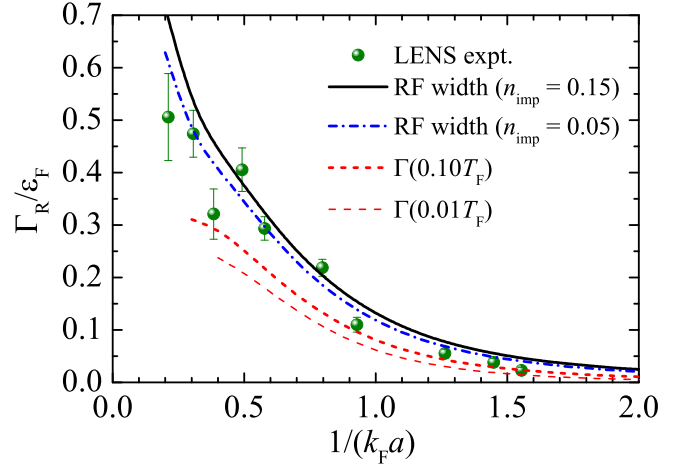}
\par\end{centering}
\caption{\label{fig11_injRFexpt} The width of the repulsive polaron extracted
from the injection rf spectrum as a function of coupling constant
$1/(k_{F}a)$ at two impurity densities $n_{\textrm{imp}}=0.15$ (black
solid line) and $n_{\textrm{imp}}=0.05$ (blue dot-dashed line). The
temperature is set to be $T=0.1T_{F}$. We compare our theoretical
predictions to the experimental data measured from coherent Rabi oscillations
at $T=0.10(2)T_{F}$ and $n_{\textrm{imp}}=0.15(1)$ \cite{Scazza2017}.
We show also the decay rate of the zero-momentum repulsive polaron
at the temperatures $T=0.1T_{F}$ (thick red dashed line) and $T=0.01T_{F}$
(thin red dashed line).}
\end{figure}

This difference may provide a natural explanation for the discrepancy
between theory and experiment for the quasiparticle lifetime of repulsive
polarons, as recently measured at LENS by using coherent Rabi oscillations
\cite{Scazza2017}. This is shown in Fig. \ref{fig11_injRFexpt},
where the data (green circles with error bar) are compared with the
predictions from Chevy's ansatz at essentially zero temperature (thin
red dashed line at $0.01T_{F}$) and at the experimental temperature
(thick red dashed line at $0.10T_{F}$). At the interaction parameter
$1/(k_{F}a)<1$, the measured decay rate is significantly larger than
the theoretical prediction. In a recent theoretical simulation \cite{Adlong2020},
a finite-temperature variational approach has been used to simulate
the real-time dynamics of Rabi oscillations. However, the simulation
is carried out at zero momentum and hence yields a similar prediction
from the finite-temperature Chevy's ansatz (see, i.e., the thick red
dashed line at $0.10T_{F}$).

\begin{figure}
\begin{centering}
\includegraphics[width=0.48\textwidth]{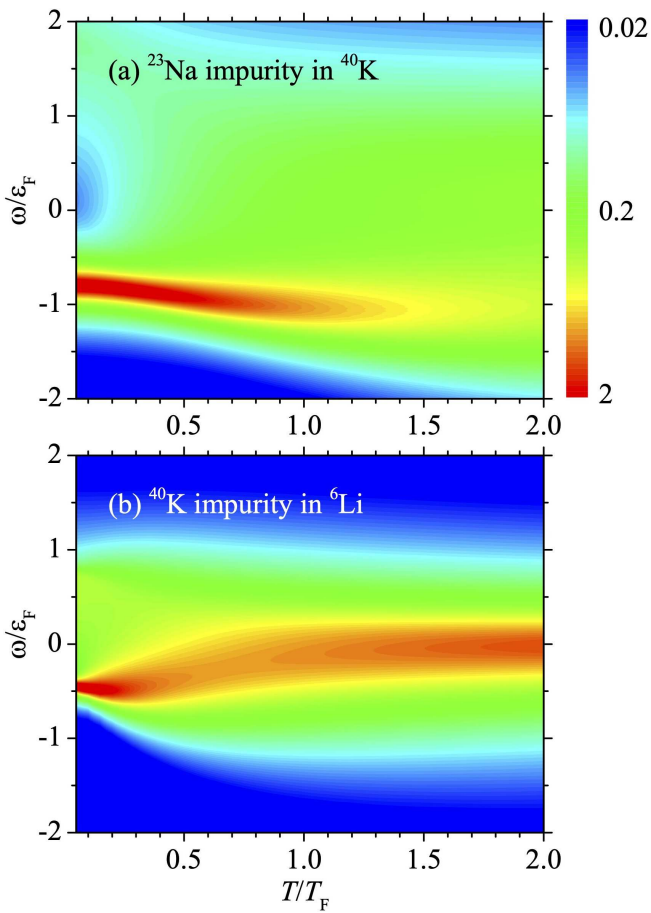}
\par\end{centering}
\caption{\label{fig12_akw2dUnequalMass} The two-dimensional contour plot of
the temperature evolution of the zero-momentum impurity spectral function
$A(\mathbf{k}=0,\omega$) for two kinds of Fermi polarons with a unitary
coupling strength $1/(k_{F}a)=0$: $^{23}$Na impurity in a Fermi
sea of $^{40}$K atoms (a) and $^{40}$K impurity in a Fermi sea of
$^{6}$Li atoms (b). Each plot is shown in the logarithmic scale.}
\end{figure}

Physically, the impurity can have a thermal distribution over different
momenta during a coherent Rabi oscillation, so we need to consider
the momentum average. Therefore, it is reasonable to assume that the
measured decay rate from Rabi oscillations might be identical to the
width measured from the injection rf spectrum. In Fig. \ref{fig11_injRFexpt},
we plot the width of repulsive polarons extracted from the theoretical
injection rf spectra, which are calculated at $T=0.1T_{F}$ and at
either the averaged experimental impurity density $n_{\textrm{imp}}\simeq0.15$
(black solid line) \cite{Scazza2017} or the minimum experimental
impurity density $n_{\textrm{imp}}\simeq0.05$ (blue dot-dashed line)
\cite{Scazza2017}. As anticipated, we find a much improved agreement
between theory and experiment, confirming the importance of the inclusion
of the momentum average.

\section{Quasiparticle lifetime at unequal mass}

\begin{figure}
\begin{centering}
\includegraphics[width=0.48\textwidth]{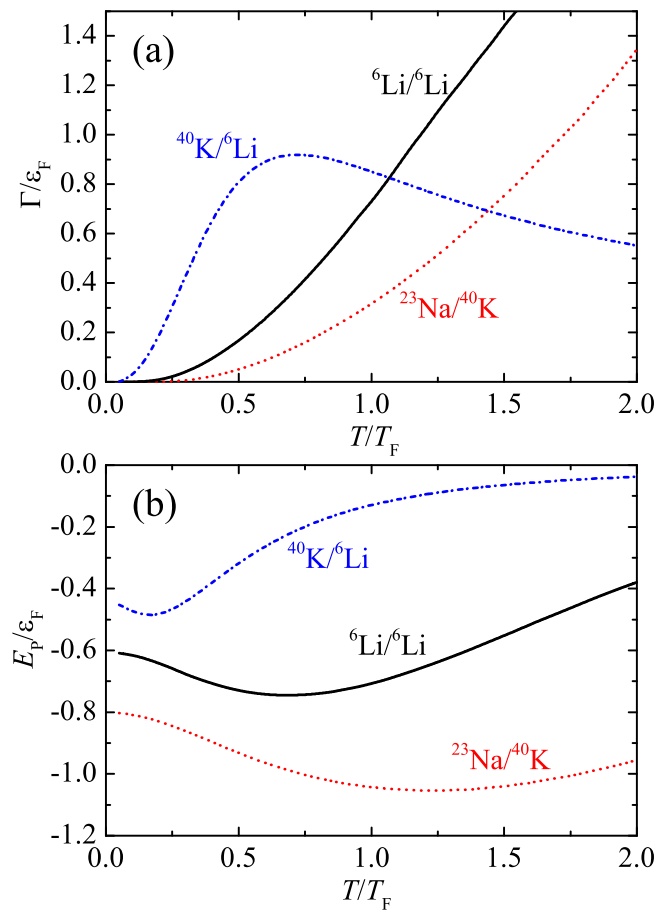}
\par\end{centering}
\caption{\label{fig13_QPUnequlMass} Quasiparticle lifetime (a) and energy
of unitary Fermi polarons as a function of temperature. Here, we consider
three experimental setups: $^{23}$Na impurity in a Fermi sea of $^{40}$K
atoms (red dotted lines), $^{6}$Li impurity in a Fermi sea of $^{6}$Li
atoms (solid lines), and $^{40}$K impurity in a Fermi sea of $^{6}$Li
atoms (blue dot-dashed lines).}
\end{figure}

We finally consider the situation that the impurity and fermionic
atoms have different masses. These cases can easily realized in experiments
by using heteronuclear atomic mixtures, such as $^{23}$Na-$^{40}$K
and $^{40}$K-$^{6}$Li mixtures. Let us focus on the most interesting
strongly-interacting limit with unitary coupling $1/(k_{F}a)=0$.

In Fig. \ref{fig12_akw2dUnequalMass}, we present the time-evolution
of the zero-momentum impurity spectral function in the form of contour
plots, for a light impurity in heavy medium (a, $^{23}$Na in a Fermi
sea of $^{40}$K atoms with $\alpha\simeq-0.27$) or a heavy impurity
in light medium (b, $^{40}$K in a Fermi sea of $^{6}$Li atoms with
$\alpha\simeq0.74$). For the former, there is no qualitative change,
in comparison with the equal mass case as reported in Fig. \ref{fig1_akw2dttf}(b).
In contrast, for the heavy impurity case, we observe a dramatic change.
As can be seen from Fig. \ref{fig12_akw2dUnequalMass}(b), by increasing
temperature the polaron peak quickly moves to zero energy. The width
of the peak initially increases with temperature. At around the Fermi
temperature, however, the width starts to become narrower. At the
largest temperature considered in the figure $T=2T_{F}$, the width
reduces to about $0.5\varepsilon_{F}$. This temperature evolution
of the width can be seen more clearly in Fig. \ref{fig13_QPUnequlMass}(a),
where we plot the decay rate as a function of temperature.

It is interesting to note that, this non-monotonic temperature dependence
of the decay rate has been previously seen in repulsive polarons close
to the Feshbach resonance, see, for example, the curve in Fig. \ref{fig4_RepulsiveDecayRT}(a)
for $1/(k_{F}a)=0.5$. Therefore, it seems reasonable to assume that
for a sufficiently heavy impurity with unitary coupling, the attractive
polaron may smoothly turn into (or more precisely, acquire the characteristic
of) a repulsive polaron at large temperature. It is also worth noting
that the temperature dependence of the decay rate and the polaron
energy of the $^{40}$K/$^{6}$Li case is qualitatively similar to
what we have observed in the experimental data for the width and peak
position of the rf spectroscopy \cite{Zan2019}, as reported in Fig.
\ref{fig8_RFexpt2}, although the latter is for the equal mass case
(i.e., $^{6}$Li/$^{6}$Li) and there is a momentum-average in the
ejection rf spectrum, as we frequently emphasized.

\section{Conclusions and outlooks}

In summary, we have presented a systematic study of finite-temperature
quasiparticle properties of Fermi polarons, by using the well-established
non-self-consistent many-body $T$-matrix theory \cite{Combescot2007}.
Different from the previous works \cite{Mulkerin2019,Tajima2019},
we have focus on the single-impurity limit, and have accurately calculated
the impurity spectral function and the associated ejection and injection
radio-frequency (rf) spectroscopies, by avoiding the ill-defined numerical
analytic continuation. Our non-self-consistent $T$-matrix calculations
also complement the earlier theoretical investigations based on a
finite-temperature variational approach \cite{Liu2019,Liu2020}.

One key result of this work is that we have clarified the important
role played by the momentum-average, which is unavoidable in the current
rf spectroscopy. As a result, the experimentally measured peak position
and width from the rf spectroscopy do not exactly correspond to the
polaron energy and decay rate. In particular, the measured width can
differ significantly from the decay rate of Fermi polarons at zero
momentum that we want to determine. By taking into account the crucial
role of momentum-average, we have successfully explained the measured
ejection rf spectrum of a unitary Fermi polaron at low temperatures
(i.e., $T<0.5T_{F}$) from the MIT group \cite{Zan2019}. We have
also resolved a puzzling discrepancy between theory and experiment
for the quasiparticle lifetime of repulsive polarons, observed in
a recent experiment at LENS \cite{Scazza2017,Adlong2020}.

The non-self-consistent $T$-matrix theory seems to work very well
for weak-coupling Fermi polarons (i.e., attractive polarons on the
BCS side and repulsive polarons on the BEC side). In the strong-coupling
unitary limit, the comparison between the theory and the MIT experiment
indicates that the theory also provides a satisfactory description
of attractive Fermi polarons at arbitrary temperatures. However, the
theory seems to strongly underestimates the precursor of repulsive
polarons near and above the Fermi degenerate temperature. We believe
this is due to the inadequate description of the vertex function,
which plays the role of the molecule Green function. In future studies
for an improved theory, it will be useful to consider the self-energy
renormalization and vertex renormalization \cite{Roulet1969}.
\begin{acknowledgments}
This research was supported by the Australian Research Council's (ARC)
Discovery Program, Grant No. DP180102018 (X.-J.L).
\end{acknowledgments}

\appendix
%dummy comment inserted by tex2lyx to ensure that this paragraph is not empty

\section{The many-body part of the two-particle propagator $\chi_{R}^{(mb)}$}

In Eq. (\ref{eq:kappamb}), let us introduce a new variable $\mathbf{p}=\mathbf{k}+(1-\alpha)\mathbf{q}/2$,
and rewrite the expression into the form, 
\begin{equation}
\chi_{R}^{(mb)}=\frac{1+\alpha}{4\pi^{2}}\intop_{0}^{\infty}\frac{dpg\left(p\right)p^{2}}{\frac{1+\alpha}{2}\left[\Omega^{+}+\mu-\frac{\left(1-\alpha\right)}{2}q^{2}\right]-p^{2}},\label{eq:kappambA1}
\end{equation}
where we have defined an angle-integrated function ($z=e^{\beta\mu}$
is the fugacity),
\begin{align}
g\left(p\right) & \equiv\intop_{-1}^{+1}\frac{dx}{2}f\left[p^{2}+\frac{\left(1-\alpha\right)^{2}}{4}q^{2}+\left(1-\alpha\right)pqx-\mu\right]\nonumber \\
 & =\frac{1}{2\left(1-\alpha\right)\beta pq}\ln\left(\frac{1+ze^{-\beta\left[p-\frac{1-\alpha}{2}q\right]^{2}}}{1+ze^{-\beta\left[p+\frac{1-\alpha}{2}q\right]^{2}}}\right).
\end{align}
The integral Eq. (\ref{eq:kappambA1}) is well defined if 
\[
-x^{2}\equiv\frac{1+\alpha}{2}\left[\Omega+\mu-\frac{\left(1-\alpha\right)}{2}q^{2}\right]<0.
\]
In this case, we find that,
\begin{eqnarray}
\textrm{Re}\chi_{R}^{(mb)} & = & -\frac{1+\alpha}{4\pi^{2}}\intop_{0}^{\infty}dp\frac{p^{2}}{p^{2}+x^{2}}g\left(p\right),\\
\textrm{Im}\chi_{R}^{(mb)} & = & 0.
\end{eqnarray}
Otherwise, let us define 
\[
y\equiv\frac{1+\alpha}{2}\left[\Omega+\mu-\frac{\left(1-\alpha\right)}{2}q^{2}\right]\geq0
\]
and use the identity (the notation P.V. means taking Cauchy principle
value)
\begin{equation}
\frac{1}{X+i0^{+}}=\mathcal{\textrm{P.V.}}\left(\frac{1}{X}\right)-i\pi\delta\left(X\right)
\end{equation}
to rewrite the real and imaginary parts of $\chi_{R}^{(mb)}$:
\begin{align}
\textrm{Re}\chi_{R}^{(mb)} & =-\frac{1+\alpha}{4\pi^{2}}\textrm{P.V.}\intop_{0}^{\infty}dp\frac{p^{2}g\left(p\right)}{p^{2}-y}\nonumber \\
 & =-\frac{1+\alpha}{8\pi^{2}}\left(C_{1}+C_{2}\right)
\end{align}
and 
\begin{align}
\textrm{Im}\chi_{R}^{(mb)} & =\frac{1+\alpha}{4\pi^{2}}\left(-\pi\right)\intop_{0}^{\infty}dpg\left(p\right)p^{2}\delta\left(y-p^{2}\right)\nonumber \\
 & =-\frac{1+\alpha}{8\pi}\sqrt{y}g\left(\sqrt{y}\right).
\end{align}
Here, by taking Cauchy principle value we have defined two integrals,
\begin{eqnarray*}
C_{1} & \equiv & \intop_{y}^{\infty}\frac{d\xi}{\xi}\sqrt{y+\xi}g\left(\sqrt{y+\xi}\right),\\
C_{2} & \equiv & \intop_{0}^{y}d\xi\frac{\sqrt{y+\xi}g\left(\sqrt{y+\xi}\right)-\sqrt{y-\xi}g\left(\sqrt{y-\xi}\right)}{\xi}.
\end{eqnarray*}

The numerical calculation of $\textrm{Re}\chi_{R}^{(mb)}$ and $\textrm{Im}\chi_{R}^{(mb)}$
therefore involves only the one-dimensional integral, which is very
efficient to carry out.

\begin{figure}
\begin{centering}
\includegraphics[width=0.48\textwidth]{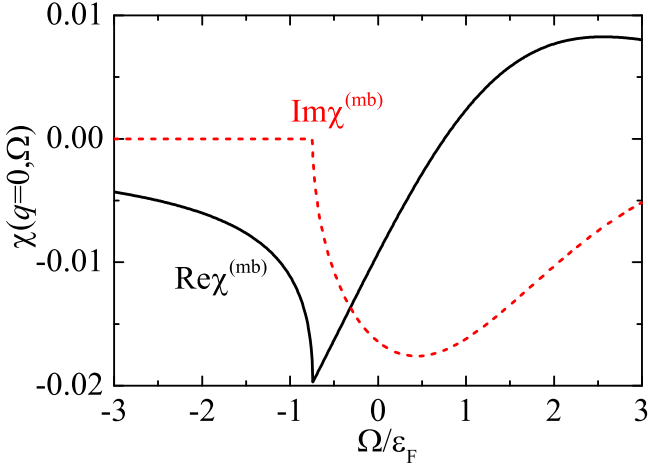}
\par\end{centering}
\caption{\label{figA1_kappa} The real and imaginary parts of $\chi_{R}^{(mb)}(\mathbf{q}=0,\Omega)$,
in arbitrary units, at the temperature $T=0.5T_{F}$. }
\end{figure}

In Fig. \ref{figA1_kappa}, we show $\chi_{R}^{(mb)}(\mathbf{q}=0,\Omega)$
at a typical temperature $T=0.5T_{F}$. The imaginary part becomes
nonzero once the frequency is above the threshold $\Omega_{c}(\mathbf{q})=(1-\alpha)q^{2}/2-\mu$,
where the chemical potential $\mu(T)$ is the temperature dependent.
As a result, there is a sharp kink in the real part $\textrm{Re}\chi_{R}^{(mb)}$. 

In general, $\textrm{Re}\chi_{R}<0$ for $\Omega<\Omega_{c}(\mathbf{q})$.
However, with $1/(k_{F}a)>0$ we may have $\textrm{Re}\chi_{R}=0$
at a certain value $\Omega<\Omega_{c}(\mathbf{q})$ and hence a pole
in the two-particle vertex function. This indicates the formation
of an undamped molecule excitation below the two-particle threshold.


\begin{thebibliography}{10}
\bibitem{NoteImpuritymu}We note that, in the spectral function $A(\mathbf{k},\omega$)
the energy $\omega$ is measured with respect to the impurity chemical
potential $\mu_{I}$. This is convenient in the single impurity limit,
where it is meaningless to specify $\mu_{I}$.
\end{thebibliography}
\end{document}